\documentstyle[multicol,aps,psfig]{revtex}
\renewcommand{\narrowtext}{\begin{multicols}{2}
\global\columnwidth20.5pc\noindent}
\renewcommand{\widetext}{\end{multicols}
\global\columnwidth42.5pc}
\begin{document}
\draft
\preprint{March 1999}
\title{Metamagnetism of antiferromagnetic $XXZ$ quantum spin chains}
\author{T$\hat{\mbox o}$ru Sakai}
\address
{Faculty of Science, Himeji Institute of Technology,
 Ako, Hyogo 678-1297, Japan}
\author{Minoru Takahashi}
\address
{Institute for Solid State Physics, University of Tokyo,
 Roppongi, Minato-ku, Tokyo 106-8666, Japan}
\date{March 1999}
\maketitle
\begin{abstract}
The magnetization process of the one-dimensional antiferromagnetic  
Heisenberg model with the Ising-like anisotropic exchange interaction 
is studied by the exact diagonalization technique. 
It results in the evidence of the first-order spin flop transition 
with a finite magnetization jump in the N\'eel ordered phase 
for $S\geq 1$. 
It implies that the $S={1\over 2}$ chain is an exceptional case 
where the metamagnetic transition becomes second-order due to 
large quantum fluctuations. 
\end{abstract}
\narrowtext

Metamagnetism is one of interesting topics in the field of the
magnetism. The spin flopping of the antiferromagnets 
with the Ising-like anisotropic exchange interaction is a simple 
mechanism of the field-induced metamagnetic transition.\cite{neel} 
In the process the spins abruptly changes directions from parallel 
to perpendicular with respect to the easy axis of the sublattice 
magnetization at some critical magnetic field. 
In the classical spin systems it is easily understood that  
the transition is first-order and accompanied by a finite magnetization
jump independently of the lattice dimension.   
In the quantum systems, however, the process is not so trivial 
because the sublattice magnetization shrinks or vanishes due to 
quantum fluctuation even at zero temperature. 
Such quantum effect is expected to be larger in lower dimensions. 
In one dimension (1D) the $S={1\over 2}$ 
Ising-like $XXZ$ model was proved by the 
Bethe ansatz exact solution\cite{yang} 
to exhibit a second-order metamagnetic 
transition at some critical magnetic field. 
In contrast the recent numerical analysis\cite{kohno} suggested that the
correponding transition is first-order for the same model in 
two and three dimensions (2D, 3D), while the magnetization jump has a 
quantum spin reduction. 
It implies that the quantum fluctuation plays such an important role 
as to change the order of the phase transition in low dimension. 
As far as we restrict us on the $S={1\over 2}$ model, 
the critical dimension seems to lie between one and two.  
On the other hand, 
the strength of the quantization also depends on the spin value $S$. 
Since the quantum effect is smaller for larger $S$, 
the second-order transition descovered for the $S={1\over 2}$ chain 
is not necessarily a common feature for all the values of $S$. 
At least the transition is first-order in the infinite $S$ limit.
Thus it is important to investigate the critical value $S_c$ between 
the second- and first-order transitions, for understanding the quantum
effect on the metamagnetism in 1D. 
In this paper, 
we study the magnetization process of the $S=1$ and ${3\over 2}$ $XXZ$
spin chains, particularly on the possibility of the first-order 
spin flopping transition. 

\begin{figure}
\mbox{\psfig{figure=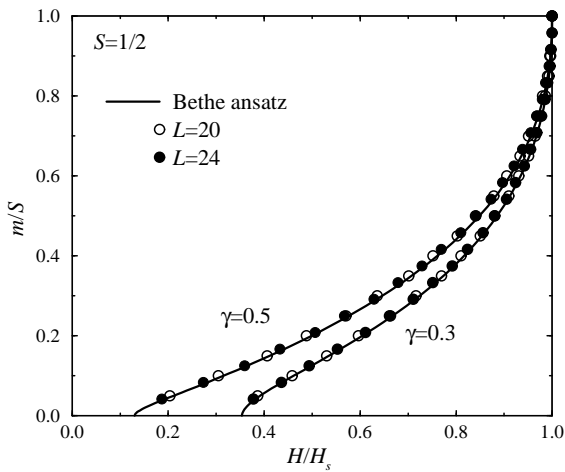,width=85mm,angle=-90}}
\vskip 5mm
\caption{
Magnetization curve for $S={1\over 2}$ obtained by the Bethe ansatz 
(solid lines), and exact diagonalization of the finite chains 
with $L=20$ and 24 (symbols). 
}
\label{fig1}
\end{figure}

\begin{figure}
\mbox{\psfig{figure=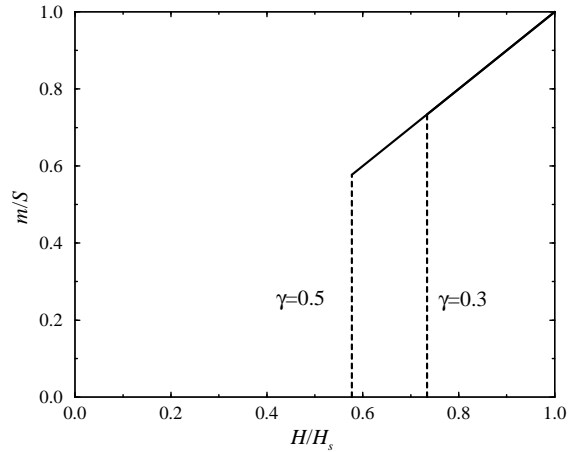,width=85mm,angle=-90}}
\vskip 5mm
\caption{
Magnetization curves in the classical limit. 
}
\label{fig2}
\end{figure}

The magnetization process of the 1D $XXZ$ model is described by the 
Hamiltonian  
\begin{eqnarray}
\label{ham}
&{\cal H}&={\cal H}_0+{\cal H}_Z, \nonumber \\
&{\cal H}_0& = \sum _j \gamma (S^x_j S^x_{j+1}+S^y_jS^y_{j+1})
+S^z_jS^z_{j+1} , \\
&{\cal H}_Z& =-H\sum _j S_j^z, \nonumber
\end{eqnarray}
under the periodic boundary condition.
We restrict us on the Ising-like anisotropy $0<\gamma <1$ to 
consider the metamagnetic transition from the N\'eel ordered 
ground state, induced by the external magnetic field along the 
easy axis of the sublattice magnetization. 
In the $S={1\over 2}$ chain a second-order transition occurs 
at some critical field $H_c$ for $0< \gamma<1$. 
The asymptotic behavior has the form 
\begin{equation}
   m\sim (H-H_c)^{1/2 }.
   \label{hc}
\end{equation}
The magnetization curves in the thermodynamic limit 
for $\gamma=$0.3 and 0.5 calculated by the Bethe ansatz are shown 
as solid lines in Fig. \ref{fig1}, where $H_s$ is the saturation field
($H_s=2S(1+\gamma)$).  
On the other hand, 
in the limit of $S\rightarrow \infty$ we know that 
the spin flop occurs at $H_c=H_s[(1-\gamma)/(1+\gamma)]^{1/2}$. 
This is the first-order transition and the magnetization jumps from 
0 to $S[(1-\gamma)/(1+\gamma)]^{1/2}$, as shown in Fig. \ref{fig2}. 
In order to investigate the corresponding transition for finite $S$ 
larger than 1/2, 
we perform the exact diagonalization technique applied to the finite 
clusters with the system size $L$, 
because there is no exact solution available here. 
Using the Lanczos algorithm, we calculate 
the lowest eigenvalue of ${\cal H}_0$ 
in the subspace with $\sum _jS^z_j=M$ for the $L$-site system, 
defined as $E(L,M)$. 
To derive the magnetization curve, 
the magnetic field $H$ to bring about the magnetization $m={M\over L}$
and $m'=m+{1\over 2}$ is estimated by $[E(L,M+1)-E(L,M-1)]/2$ and 
$E(L,M+1)-E(L,M)$, respectively. 
To demonstrate the validity of the method, 
we show the results of the $S={1\over 2}$ chain by the diagonalization
for $L=20$ and 24 as circles, together with the Bethe ansatz solution in Fig.
\ref{fig1}. 
The bulk property is revealed to be well understood even in such small systems.

In the small magnetization limit in the N\'eel ordered phase, 
only the state with $M=2Sn$ $(n=0,1,2,\cdots)$ can contribute the 
ground-state magnetization process of the spin-$S$ chain. 
Because states where every site has $S^z_j=+S$ or $-S$ are stabilized 
better than other states.
To explain the reason, we consider the $S=1$ chain, for example. 
In the excited state produced by changing $S^z_j$ from $-1$ to 0 
in the N\'eel state ($\cdots \downarrow \uparrow 0 \uparrow \downarrow
\cdots $), the motion of the object 0 to the adjacent site 
($\cdots \downarrow \uparrow \uparrow 0 \downarrow \cdots $) 
will lead to some energy loss in the diagonal element of the exchange 
Hamiltonian, while it will gain some energy in the off-diagonal element.
In contrast 
the object $\uparrow$ doped in the N\'eel order 
($\cdots \downarrow \uparrow \uparrow \uparrow \downarrow \uparrow \cdots $) 
can yield the energy gain due to the motion along the chain 
with no loss in the diagonal element 
($\cdots \downarrow \uparrow \uparrow \downarrow \uparrow \uparrow \cdots $). 
Thus smaller magnetic field can excite $\downarrow$ to $\uparrow$, 
rather than to 0. 
It implies that states with $+1$ or $-1$ at every site are more favorable 
than states including 0, and the magnetization process starting from the
N\'eel phase consists of the states with $M=2n$ in the small $m$ limit.
The argument is also generalized for arbitrary $S$ and 
the state with $M=2Sn$ is stabilized. 
Thus in the small magnetization region an oscillation 
with the period of $2S$ in the $M$ dependence of the energy of 
the spin-$S$ chain in the Ising phase.   
To avoid the oscillation, 
we will take only states with $M=2Sn$ into account in the following  
analysis, 
except for the determination of the saturation field 
$H_s$.   
\begin{figure}
\mbox{\psfig{figure=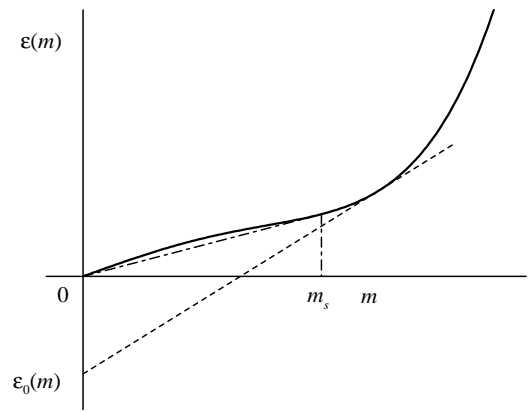,width=85mm,angle=-90}}
\vskip 5mm
\caption{
Schematic curve of $\epsilon (m)$ in the case when the first-order 
metamagnetic transition occurs. $m_s$ is the magnetization jump and 
it is determined by $\epsilon _0(m)=0$. 
}
\label{fig3}
\end{figure}

\begin{figure}
\mbox{\psfig{figure=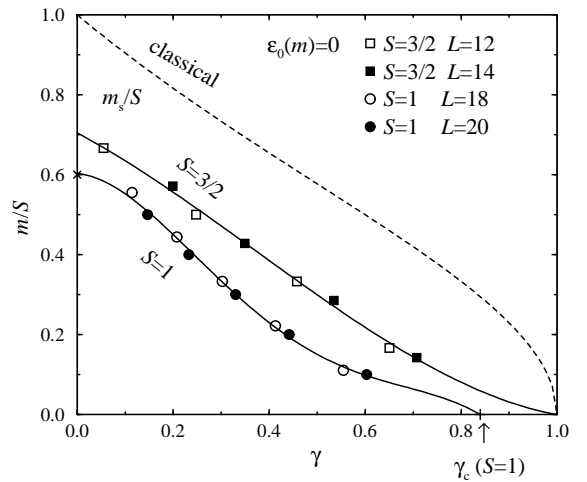,width=85mm,angle=-90}}
\vskip 5mm
\caption{
Magnetization jump $m_s/S$ for $S=1$ and ${3\over 2}$ (solid lines) 
obtained by the polynomial fitting to the points with $\epsilon _(0)=0$
for the finite systems (symbols). For $S=1$ $\gamma _c$ is the boundary
between the N\'eel and Haldane phases, and $\times$ is the upper bound
of $m_{s0}$. 
$m_s/S$ in the classical limit is a dashed line. 
}
\label{fig4}
\end{figure}

In general, a first-order metamagnetic transition occurs, 
when the ground state energy per site $\epsilon (m)$ for the Hamiltonian
${\cal H}_0$ (We choose the origin to set $\epsilon (0)=0$ here.) 
has such $m$-dependence as shown in Fig. \ref{fig3}, 
where the region with $\epsilon ''(m)< 0$ lies from the origin 
to some magnetization. 
We define $\epsilon _0(m)$ as the intersection of the vertical axis and
the tangent line of the curve $\epsilon (m)$, as shown in Fig.
\ref{fig3}. 
The magnetization jump $m_s$ is determined as the solution of 
$\epsilon _0(m)=0$. 
It implies that the sign of $\epsilon _0(m)$ changes at $m_s$ 
and the magnetization with $\epsilon _0(m)>0$ does not appear 
in the ground-state magnetization process. 
In order to examine the possibility of the magnetization jump 
in the spin-$S$ chains, 
we investigate the behavior of $\epsilon _0(m)$  
based on the finite cluster calculation.  
We assume the size dependence of the ground state energy per site 
obeys the relation 
\begin{eqnarray}
\label{gse}
e(L,M)\equiv 
{1\over L}[E(L,M)-E(L,0)] \sim \epsilon (m) + O({1\over {L^2}}),
\end{eqnarray}  
for $L\rightarrow \infty$ with fixed $m={M\over L}$. 
Although the relation is predicted by the conformal field theory
\cite{cft} for massless state, 
the following argument will be valid even in massive cases,  
as far as the size correction does not exceed $O({1\over {L^2}})$ 
We also define $h(L,M)$ as 
\begin{eqnarray}
\label{h}
h(L,M)&\equiv& {1\over {2S}}[E(L,M+2S)-E(L,M-2S)] \nonumber \\ 
&\sim& \epsilon '(m)+O({1\over {L^2}}), 
\end{eqnarray}  
where the size dependence is derived from (\ref{gse}). 
The quantity $h(L,M)$ should converge to $H$ for $L\rightarrow \infty$
in the normal magnetization process. 
Since $\epsilon _0(m)$ is given by 
$\epsilon _0(m)= \epsilon (m)-\epsilon '(m)m$, 
it can be estimated by the form 
\begin{eqnarray}
\label{e0}
{1\over L}[e(L,M)-h(L,M)M]\sim \epsilon _0(m)+O({1\over {L^2}}). 
\end{eqnarray}  
Since various system sizes with fixed $m={M\over L}$ are available 
only for few values of $m$ because of the restriction $M=2Sn$, 
we neglect the size correction in (\ref{e0}) in the following analysis.
For $S={1\over 2}$ the finite cluster calculation up to $L=26$ 
indicates $\epsilon _0 (m) <0$ for $0<m<1$ 
in all the region $0<\gamma <1$.  
It means no first-order transition, which is consistent wiht the 
result from the Bethe ansatz. 
For $S=1$, however, the calculation up to $L=20$ detects 
the region with $\epsilon _0(m)>0$, which suggests the first-order 
metamagnetic transition.  
When we vary $\gamma$ with fixed $M$, 
a point with $\epsilon _0(m)=0$ can be found for small $m$. 
The points for various values of $M$ available for $L=18$ and 20 
are plotted on the $m/S$-$\gamma$ plane as circles in Fig. \ref{fig4}. 
Since $\epsilon _0(m)$ is positive under the points, while negative 
over them in the plane, 
the points stand for the magnetization jump $m_s$ for corresponding $\gamma$.
In the analysis of 20-site cluster, 
$\epsilon _0(m)$ is always positive for $M\geq 12$. 
Thus we think that $m=0.6$ is an upper bound of $m_s$ in the 
limit $\gamma \rightarrow 0$ denoted as $m_{s0}$, which is shown as 
$\times$ in Fig. \ref{fig4}. 
$m_{s0}$ does not correspond to the value of the Ising model 
$m_s/S=1$, which implies that there exists a spin reduction 
due to the quantum fluctuation, as was found in the 2D and 3D 
$S={1\over 2}$ model.\cite{kohno} 
On the other hand, 
the $S=1$ $XXZ$ chain has the Haldane phase\cite{haldane}, 
where the second-order transition described by (\ref{hc}) was 
revealed to occur in the magnetization process\cite{schulz,affleck,taka}, 
in the large $\gamma$ region. 
Consulting some previous works\cite{botet,nomura,gomez,sakai1}, 
we put the boundary between the Haldane and N\'eel phases at 
$\gamma _c=0.84$, as shown in Fig. \ref{fig4}. 
In the present work,  
by the analysis of several excitation gaps corresponding to
the soft modes of the Luttinger liquid, 
we checked that the magnetic state for $m_s<m<1$ consists of a 
single gapless phase characterized by the dominant spin correlation 
function $\langle S_0^+ S_r^- \rangle \sim (-1)^r r^{-\eta}$, 
in contrast to the magnetization process in the presence of the Ising-like 
single-ion anisotropy described by $D\sum _j(S^z_j)^2$ $(D<0)$, 
where there appears another massless phase\cite{schulz, sakai} 
with $\langle S^z_0S^z_r\rangle \sim \cos(2k_Fr)r^{-\eta '}$.  
Thus the line of $m_s$ is expected to continue to the unique phase 
boundary $\gamma _c$ at $m=0$. 
The solid line for $S=1$ in Fig. \ref{fig4} is the result from the 
polynomial fitting to the points with $\epsilon _0(m)=0$ ($L=18$ and 20) 
and the boundary $\gamma _c$, also adding $m_{s0}=0.6$ because 
the fitted line excluding it would exceed the upper bound.  
The points with $\epsilon _0(m)=0$ for $S={3\over 2}$ are also plotted
as squares in Fig. \ref{fig4}, 
where we use the results of $L=12$ and 14. 
Assuming the boundary of the N\'eel phase lies at $\gamma =1$, 
the curve fitted to the points is shown as a solid line in Fig.
\ref{fig4}. 
In this time 
$m_{s0}$ is estimated from the fitted curve as $m_{s0}=0.70 \pm 0.04$,
because it is smaller than the upper bound determined 
by $m$ for which $\epsilon _0(m)$ 
is always negative. 
The spin reduction in the limit $\gamma \rightarrow 0$ is smaller 
than that for $S=1$. 
It is consistent with the reasonable assumption that 
the quantum fluctuation is smaller for larger $S$. 
The line of the magnetization jump in the classical limit 
$m_s/S=[(1-\gamma )/(1+\gamma )]^{1/2}$ is also plotted as a dashed curve 
in Fig. \ref{fig4}. 
These curves suggest that the magnetization jump increases with 
the spin value increasing from $S=1$ towards the classical limit. 
Therefore we conclude that the 1D Ising-like $XXZ$ model 
($0<\gamma <1$) exhibits  
the first-order spin flopping transition for $S\geq 1$, 
except for the Haldane phase.  
It implies that the $S={1\over 2}$ chain is an exceptional case 
where the metamagnetic transition from the N\'eel phase is 
second-order, 
that is, the critical spin value $S_c$ lies between ${1\over 2}$ and 1.  
\begin{figure}
\mbox{\psfig{figure=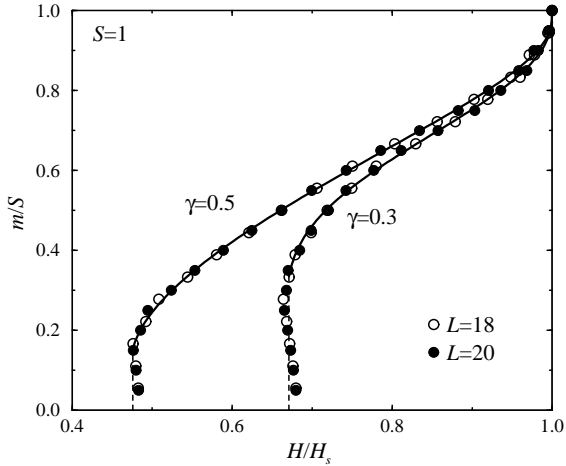,width=85mm,angle=-90}}
\vskip 5mm
\caption{
Magnetization curves for $S=1$. 
}
\label{fig5}
\end{figure}
\begin{figure}
\mbox{\psfig{figure=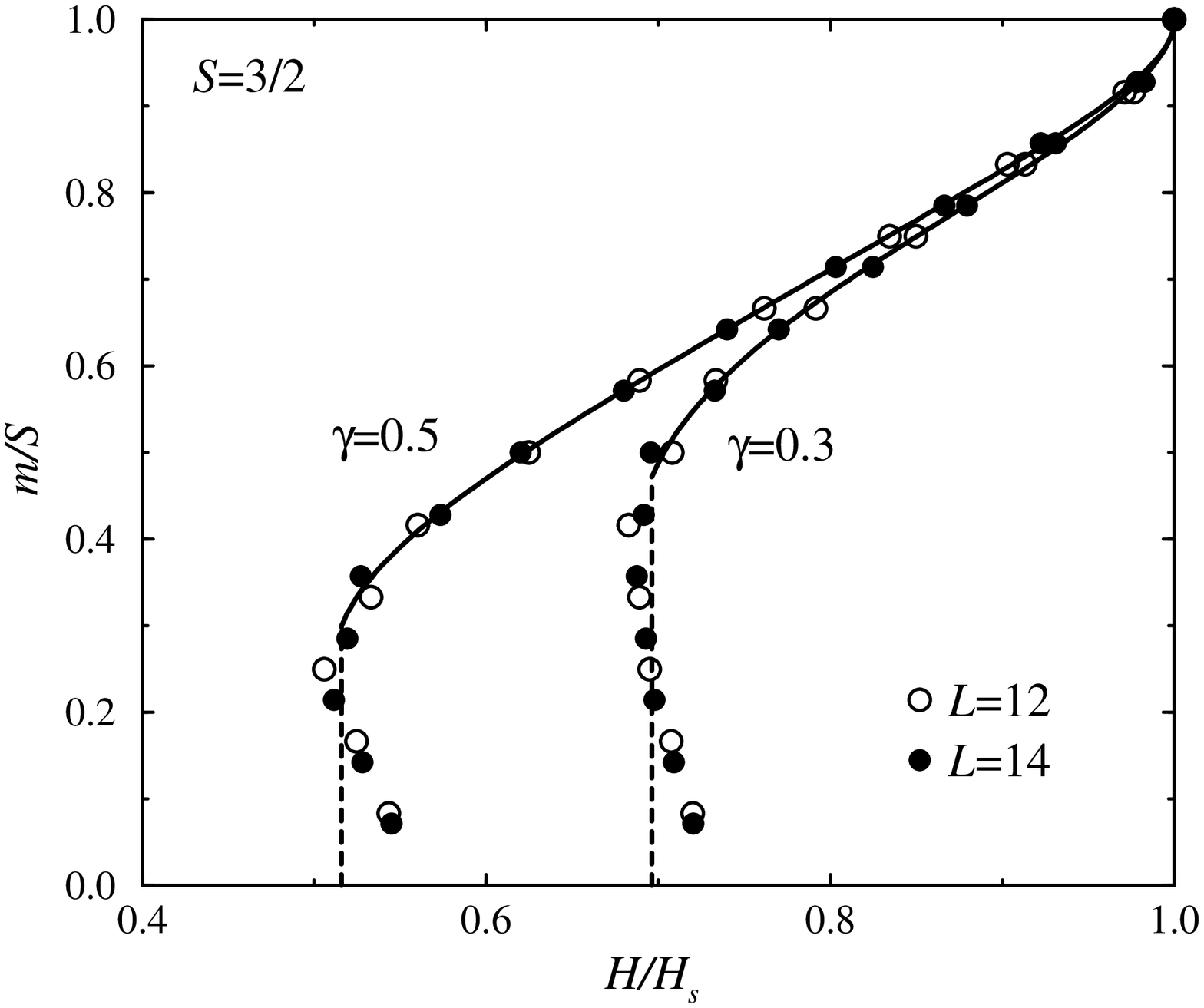,width=85mm,angle=-90}}
\vskip 5mm
\caption{
Magnetization curves for $S={3\over 2}$. 
}
\label{fig6}
\end{figure}

In order to convince of the conclusion, 
we show the ground-state magnetization curves  
for $S=1$ and ${3\over 2}$ based on the present finite cluster 
calculation. 
We plot $h(L,M)$ as the field $H$ for the magnetization $m={M\over L}$, 
and $h'(L,M)\equiv [E(L,M+2S)-E(L,M)]/(2S)$ for $m=(M+S)/L$, 
using the value renormalized by the saturation field $H_s$. 
The results of the largest two system sizes 
for $S=1$ and ${3\over 2}$ ($\gamma =0.3$ and 0.5) 
are shown as symbols in Figs. \ref{fig5} and \ref{fig6}, respectively. 
They all indicate the evidence of the first-order transition, 
in contrast to the $S={1\over 2}$ in Fig. \ref{fig1}.  
We determine 
the magnetization jump $m_s$ from the fitted lines in Fig. \ref{fig4},
and put the fitted curves for $m_s <m$ in Figs. \ref{fig5} and
\ref{fig6}, to complete the magnetization curves. 
The behaviors of the symbols in Figs. \ref{fig5} and \ref{fig6} 
suggest that the size correction is so small that the magnetization 
curves given by the solid lines  are expected to be of the bulk systems.

The present analysis indicated a numerical evidence of the 
first-order metamagnetic transition in the 1D $XXZ$ model 
for $S\geq 1$, based on the finite-cluster calculation. 
In general the finite system has larger quantum 
fluctuation, which tends to suppress the magnetization jump,  
than the infinite system. 
Thus the present conclusion on the finite jump is expected to be 
valid even in the thermodynamic limit.  

In summary, 
the exact diagonalization study on the finite cluster 
suggested that the 1D Ising-like antiferromagnetic $XXZ$ model 
exhibits the first-order spin flopping transition with a finite 
magnetization jump for $S\geq 1$, except for the Haldane phase. 
It implies that 
only for $S={1\over 2}$ the quantum fluctuation is extremely large 
and makes the corresponding transition second-order. 

The numerical computation was done using the facility of the
Supercomputer Center, Institute for Solid State Physics, University of
Tokyo.

\widetext
\end{document}